\documentstyle[prl,aps,preprint]{revtex}
\begin{document}
\draft
\title{Coexistence of Antiferromagnetism and Superconductivity near the Quantum Criticality in Heavy Fermion Compound CeRhIn$_5$}
\author{T.~Mito$^{1,*}$, S.~Kawasaki$^1$, Y.~Kawasaki$^1$, G.~-q.~Zheng$^1$, Y.~Kitaoka$^1$, D. Aoki$^{2,**}$, Y.~Haga$^3$,  and Y.~\=Onuki$^2$}
\address{$^1$Department of Physical Science, Graduate School of Engineering Science, Osaka University, Toyonaka, Osaka 
560-8531, Japan}
\address{$^2$Department of Physics, Graduate School of  Science, Osaka University, Toyonaka, Osaka 560-0043, Japan}
\address{$^3$Advanced Science Research Center, Japan Atomic Energy Research Institute, Tokai, Ibaraki 319-1195, Japan}
\date{\today}
\maketitle 
\widetext
\begin{abstract}
We report a study on the interplay between antiferromagnetism (AFM) and superconductivity (SC) in a heavy-fermion compound CeRhIn$_5$ under pressure $P=1.75$ GPa. The onset of the magnetic order is evidenced from a clear split of $^{115}$In-NQR spectrum due to the spontaneous internal field below the N\'eel temperature $T_N=2.5$ K. Simultaneously, bulk SC below $T_c=2.0$ K is demonstrated by the observation of the Meissner diamagnetism signal whose size is the same as in the exclusively superconducting phase. These results indicate that the AFM coexists homogeneously with the SC at a microscopic level.
\end{abstract}

\pacs{PACS: 74.25.Ha, 74.62.Fj, 74.70.Tx, 75.30.Kz, 76.60.Gv}

\newpage

\narrowtext
Understanding the interplay between magnetism and superconductivity (SC) is one of the central tasks in condensed-matter physics.
In many strongly-correlated electron systems such as high transition-temperature ($T_c$) copper oxides, low-dimensional organic salts and heavy-mass electron (heavy fermion) intermetallic compounds, the magnetism and SC are closely related.
In some of heavy-fermion (HF) systems, the relationship between the two phenomena are complicated; antiferromagnetically-ordered and superconducting phases appear to coexist in some cases, which has been reported particularly in some uranium (U)-based HF systems.
For example, UPd$_2$Al$_3$ exhibits a superconducting transition at $T_c=1.8$ K well below an antiferromagnetic ordering temperature $T_N=14.3$ K \cite{Geibel,Steglich}.
The U-based HF systems generally involve more than two $5f$ electrons.
In UPd$_2$Al$_3$, it is suggested that two of them, that possess a localized character, are responsible for the antiferromagnetism (AFM), whereas the remaining $5f$ electrons, which are hybridized with conduction electrons and become heavy, are responsible for the SC.
Sato {\it et al.} have proposed a picture that magnetic excitons resulting from the exchange interactions between magnetic moments produce effective interactions between itinerant electrons and are responsible for the onset of the SC \cite{Sato}.
Quite recently, the SC was discovered even in the ferromagnetic materials UGe$_2$ and URhGe \cite{Saxena,Aoki}.
Although the detailed mechanism for the SC has not been identified yet in these materials \cite{Flouquet}, the fact that the SC and ferromagnetism disappear at a same value of critical pressure in UGe$_2$ gives a hint for an intimate relationship between the two types of orderings \cite{Saxena,Tateiwa}.

In contrast to these U-based HF materials, Ce-based HF compounds involve only one 4$f$ electron per Ce ion, which may participate in both magnetism and SC.
Measurements of resistivity under pressure ($P$) in CeCu$_2$Ge$_2$, CePd$_2$Si$_2$ and CeIn$_3$ also suggest a coexistence of the AFM and SC in a narrow $P$ range \cite{jaccard92,Mathur,Grosche}.
It has, however, still remained controversial whether the coexistence of the two phases is homogeneous or they are spatially segregated.

The finding of a new Ce-based compound CeRhIn$_5$ has opened a way to systematically investigate the evolution from the antiferromagnetic to superconducting state as a function of the pressure \cite{Hegger}.
This is because the critical pressure $P_c \sim 1.6$ GPa, at which the two phases meet, is much lower than in the previous examples.
Remarkably, its $T_c$ does not only reach a record of $T_c$=2.1 K to date, but also the SC is more robust against increasing pressure than in CePd$_2$Si$_2$ and CeIn$_3$ \cite{Hegger,Muramatsu}.
An effective moment at the paramagnetic state is estimated to be $\mu_{\rm eff}=2.38 \mu_{\rm B}$, which is somewhat reduced from a value $2.56 \mu_{\rm B}$ expected for a 4$f$ localized model \cite{Hegger}.
Below $T_N=3.8$ K, a nuclear quadrupolar resonance (NQR) study indicates the rapid development of an internal field $H_{int}$ induced by the Ce spontaneous moments and a spiral modulation of the Ce moments that is incommensurate with the lattice \cite{Curro}.
 A recent neutron experiment revealed the reduced Ce magnetic moments $0.37\mu_{\rm B}$ at 1.4 K form in a helical spiral structure along the c-axis with an incommensurate wave vector ${\rm q_M}=(1/2, 1/2, 0.297)$\cite{Bao}.
An extensive NQR study has showed that $T_N$ exhibits a moderate variation, whereas the $H_{int}$ is linearly reduced in $P=0-1.23$ GPa, extrapolated to zero at $P^*=1.6\pm 0.1$GPa \cite{Mito}.
Note that this $P^*$ is comparable to $P_c = 1.63$ GPa at which the superconducting signature appears \cite{Hegger}. The NQR experiment indicated that the $P$ induced transition from the AFM to SC is of a second-order type, but not of a first-order type that was suggeted by the measurments of resisitivity \cite{Hegger} and specific heat \cite{Fisher}. 
At $P=2.1$ GPa apart from $P_c$ or $P^*$, $1/T_1$ reveals a $T^3$ dependence below $T_c$, consistent with the existence of the line-node  gap \cite{Mito}.
This was also corroborated by the specific heat measurement \cite{Fisher}.
In alloying systems like CeRh$_{1-x}$A$_x$In$_5$ (A=Ir, Co), on the one hand, the two phases were suggested to coexist in certain concentration ranges of $x$ \cite{Pagliuso,Zapf,Zheng}.
Apparently, the nature of the $P$ induced transition from AFM to SC  has remained controversial in CeRhIn$_5$, especially near $P_c$=1.63 GPa where both the phases meet one another.

In this letter, we report results of $^{115}$In-NQR and ac-susceptibility (ac-$\chi$) experiments in CeRhIn$_5$ at $P=1.75$ GPa close to $P_c$ at zero magnetic field ($H=0$). We find an onset of the magnetic order at $T_N=2.5$ K under $P=1.75$ GPa from a clear split of NQR spectrum due to the spontaneous internal field.
Simultaneously, bulk SC below $T_c=2.0$ K is demonstrated by the observation of the Meissner diamagnetism in ac-$\chi$ whose size is the same as in the exclusively superconducting phase in $P=1.95-2.15$ GPa. We conclude that the AFM coexists homogeneously with the SC at a microscopic level.

Single crystal of CeRhIn$_5$ was grown by the self-flux method \cite{Hegger}, and was moderately crushed into grains in order to make rf pulses penetrate into samples easily.
CeRhIn$_5$ consists of alternating layers of CeIn$_3$ and RhIn$_2$ and hence has two inequivalent In sites per unit cell.
The $^{115}$In-NQR measurements were made at the In(1) site \cite{Curro,Mito,Shinji} which is located on the top and bottom faces of the tetragonal unit cell.
The NQR spectrum was obtained by plotting the intensity of spin-echo signal as a function of frequency.
The high-frequency ac-$\chi$ was carried out by measuring the inductance of an {\it in situ} NQR coil with the frequency $\sim 2.5$ MHz that is not far from the NQR frequency $\nu_Q\sim$ 7MHz. Note that the skin depth for the NQR and ac-$\chi$ measurements is comparable to a size of grains ($\sim$ 0.1mm) and the magnetic penetration depth is estimated to be typically an order of 5000-7000 $\AA$ for the superconducting CeTIn$_5$ (T=Ir, Co) \cite{Higemoto}.
So, the present ac-$\chi$ measurement is compatible to the dc-$\chi$ one in detecting a bulk SC.
The hydrostatic pressure was applied by utilizing BeCu or NiCrAl/BeCu piston-cylinder cell, filled with Daphne oil (7373) as a pressure-transmitting medium.
When using liquid as a pressure-transmitting medium to obtain quasi-hydrostatic pressure, care was taken on some inevitable pressure inhomogeneity 
in the sample.  For our pressure cells, an extent of the spatially distribution in values of pressure $\Delta P/P$ is estimated to be $\sim 3 \%$ from a broadening in the linewidth in NQR spectrum, for example, at $P$=1.75 GPa, $\Delta P\sim$0.05 GPa.

Fig. 1 indicates respective NQR spectra at  $T$=2.65 and 1.45 K at $P=1.75$ 
GPa for the 1$\nu_{Q}$ transition which is the lowest one among the four transitions for the In nuclear spin $I$=9/2. In the paramagnetic state above $T_N$, only one sharp spectrum is observed as shown in the upper panel.
As the temperature decreases, the NQR spectrum starts to broaden below $T_N=2.5$ K and eventually splits into two peaks as seen in the lower panel.
This is interpreted as follows.
When the Ce-$4f$ magnetic moments order, they induce spontaneous internal fields $H_{int}$ at the In(1) site, so that the shape of the NQR spectrum is split into two lines.
This clear split in the spectrum, which can be fitted by two Lorenztian functions as seen in the lower panel of Fig.1 (solid curve), persists down to low temperatures below $T_c$=2.0 K.
Note that the separation between the two peaks in the spectrum is directly proportional to the size of magnetically ordered moments \cite{Curro}.
The $T$ dependence of $H_{int}$ at $P=1.75$GPa is shown in the inset.
 A plot of $H_{int}$ against $T$ is fitted to $(1-T/T_N)^{\beta}$ with $\beta =0.5$ (solid line) rather than $\beta =0.25$ (broken line). The latter behavior was consistent with those at $P=0$ \cite{Curro} and at the values of lower pressures \cite{Mito}. This may suggest that the rapid development of $H_{int}$ becomes slower due to a coexistence between the SC and AFM in the vicinity of $P_c$ as demonstrated later.
Any single peak that is observed when the system is in the paramagnetic state at $T=2.65 $ K does not remain at the magnetically ordered state at $T$=1.45 K as shown in Fig.1.
This result gives direct evidence for the onset of homogeneous AFM throughout the whole sample, excluding any possibility of phase segregation at $P$=1.75 GPa near the border at which the two phases meet.
The $H_{int}$ at $P=1.75$GPa and $T=1.45$K is estimated to be $\sim 80$Oe from the separation between the two peaks, corresponding to a size of the ordered moment whose value is at most $\sim$ 5\% of the value at $P$=0 \cite{Mito}.

As seen in Fig.2, the high-frequency ac-$\chi$ measurement at $P=1.75$ GPa has revealed a clear bulk SC with  its onset temperature at $T_c=2.0 $K that is lower than $T_N=2.5$ K.
The bulk nature of SC at $P=1.75$ GPa was warranted by the observation of the same size of the Meissner diamagnetism in ac-$\chi$ in $P=1.95-2.15$ GPa. Note that at $P=2.1$ GPa the unconventional SC with the line-node gap was exclusively established from the measurements of the NQR-$T_1$ \cite{Mito} and the specific heat \cite{Fisher}. 
 The data at $P$=1.75 and 1.95 GPa are actually compared in Fig.2.
By combining this evidence for the bulk SC and the emergence of homogeneous internal field due to the AFM probed by the NQR spectrum, we conclude that the AFM and SC coexist homogeneously at a microscopic level at $P$=1.75 GPa and hence the $P$-induced transition from AFM to SC around $P_c\sim$1.63 GPa is not of any first-order type in CeRhIn$_5$, although it was suggested from the meassurements of resistivity \cite{Hegger} and specific heat \cite{Fisher}.

We summarize the $T-P$ phase diagram for CeRhIn$_5$ in Fig.3 along with the results reported previously \cite{Mito,Shinji}.
With increasing $P$, $T_N$ is suppressed at pressures exceeding $P\sim 1$ GPa. Concomitantly the low-energy spectral weight of magnetic fluctuations are suppressed. Namely, the pseudogap behavior emerges as reported in the ref.\cite{Shinji}.
Although it is not identified yet whether this behavior is relevant to  the SC or the AFM, the presence of the pseudogap behavior at the border of AFM and SC seems to share a common feature with other strongly correlated electron systems such as high-$T_c$ cuprate and layered organic superconductors \cite{Timusk,Kanoda}.  The coexistence of AFM and SC in CeRhIn$_5$, however, differs from the previous examples.
In particular, the layered organic superconductors show a first-order transition between the two phases \cite{Lefebvre}, as evidenced from the observation of two NMR spectra that arise from the antiferromagnetically ordered phase and superconducting phase near the boundary of the two phases.
This  coexistence of the AFM and SC in CeRhIn$_5$ is unconventional in the context that the HF state derived by one $4f$-electron per Ce ion  contributes to both, and hence deserves theoretical studies.
A possible scenario for this coexistence may be addressed as follows. 
When the antiferromagnetic exchange constant $J_{\rm cf}$ between $4f$ electrons and conduction electrons remains strong, $4f$ local moments are compensated due to a Kondo-like interaction and the renormalized HF state could be itinerant, leading to the occurrence of the $P$ induced SC in CeRhIn$_5$.
As $J_{\rm cf}$ decreases gradually, the HF quasiparticles tend to polarize since the Ruderman-Kittel-Kasuya-Yosida (RKKY) and the Kondo-like interactions are competing at the quantum critical region where the two phases come across.
In such the regime, however, these two effects may lead to the homogeneous coexistence of the AFM with tiny ordered moments and the SC at a microscopic level, 
instead of deserving a quantum critical point.

Finally we note that the interplay between the AFM and SC in the first Ce-based compound CeCu$_2$Si$_2$ that displays the SC at $P$=0 \cite{CeCu2Si2}, was recently discussed in the framework of the SO(5) theory \cite{kitaoka01} that unifies the AFM and SC \cite{zhang} and as a result the two phases are suggested to have a common mechanism.  It is worthwhile examining whether the coexistence of the AFM and SC in CeRhIn$_5$ can be accounted for by the SO(5) theory as well.

In conclusion, we have found via the $^{115}$ In NQR and the ac-susceptibility measurements that the AFM and SC coexist homogeneously at a microscopic level in the HF compound CeRhIn$_5$ at $P$=1.75 GPa.
It is unconventional that the Ce $4f$-electron which derives HF state contributes to both AFM and SC. This may be possible in an itinerant regime where the size of ordered moments is largely reduced as actually suggested from the experiment.  Our finding may also shed light on other strongly correlated systems including high-$T_c$ superconductors.

One of authors (T.M) thanks K. Ishida and S. Wada for their useful discussions. This work was supported by the COE Research grant (10CE2004) of Grant-in-Aid for Scientific Research from the Ministry of Education, Sport, Science and Culture of Japan.

\vspace{2cm}
\noindent
* Present address: Department of Physics, Faculty of Science, Kobe University, Nada, Kobe 657-8501, Japan
\noindent
** Present address: CEA, D\'epartment de Recherche Fondamentale sur la Mati\`ere Condens\'ee, SPSMS, 38054 Grenoble Cedex 9, France

\begin{figure}[htbp]
\caption[]{$^{115}$In-NQR spectra of 1$\nu_Q$ ($\pm 1/2 - \pm 3/2$ transition for $I=9/2$ ) at $P=1.75$GPa and $T$=2.65 K (upper panel) and 1.45 K (lower panel). In the lower panel, the dotted curves are fits by two Lorentzians which peaks are split due to the internal field $H_{int}$ associated with the onset of antiferromagnetism and the solid curve is the convolution of the two curves.
The inset shows the temperature dependence of $H_{int}$.
The solid and broken lines are fits to $(1-T/T_N)^{\beta}$ with $\beta =0.5$ and $=0.25$, respectively.
 }
\end{figure}

\begin{figure}[htbp]
\caption[]{Temperature dependence of the high-frequency ac-susceptibility measured using an {\it in-situ} NQR coil at $P=1.75$GPa (open circles) and $P=1.95$GPa (closed circles).}
\end{figure}

\begin{figure}[htbp]
\caption[]{The temperature-pressure phase diagram for CeRhIn$_5$.
The triangles, circles and crosses show $T_N$, $T_c$ and $H_{int}$, respectively, along with the previous results \cite{Mito,Shinji}. 
The closed symbols represent the present data ($P=1.75$ GPa).
The open squares, that was cited from the literature \cite{Shinji}, show the $T_{PG}$ at which $1/T_1T$ shows a peak.
The solid and broken curves are eye-guides.
The arrow points the value of pressure.
}
\end{figure}
\end{document}